\documentclass[12pt,preprint]{aastex}
\usepackage{multirow}
\usepackage{hhline}
\usepackage{url}
\usepackage{epsfig}
\usepackage{amssymb}
\usepackage{rotating}
\usepackage{hyperref}

\def\ltsima{$\; \buildrel < \over \sim \;$}
\def\lsim{\lower.5ex\hbox{\ltsima}}
\def\gsim{\lower.5ex\hbox{\gtsima}}
\def\lapp{\ifmmode\stackrel{<}{_{\sim}}\else$\stackrel{<}{_{\sim}}$\fi}
\def\gapp{\ifmmode\stackrel{>}{_{\sim}}\else$\stackrel{<}{_{\sim}}$\fi}

\def\Msun{M_{\odot}}

\newdimen\minuswidth    %define @ width of minus sign for tables

\setbox0=\hbox{$-$}

\minuswidth=\wd0

\catcode`@=\active

\def@{\kern\minuswidth}

\setbox0=\hbox{\rm0}
\shorttitle{MSP companions in 47 Tucanae}

\shortauthors{}

\begin{document} 

\title{Optical Identification of He White Dwarfs Orbiting Four
  Millisecond Pulsars in the Globular Cluster 47 Tucanae\footnote
  {Based on observations collected with the NASA/ESA HST
    (Prop. 12950), obtained at the Space Telescope Science Institute,
    which is operated by AURA, Inc., under NASA contract NAS5-26555.}}

\author{ M. Cadelano\altaffilmark{1,2}, C. Pallanca\altaffilmark{1},
  F. R. Ferraro\altaffilmark{1}, M. Salaris\altaffilmark{3},
  E. Dalessandro\altaffilmark{1},
  B.~Lanzoni\altaffilmark{1},~P.~C.~C.~Freire\altaffilmark{4}}
\affil{\altaffilmark{1} Dipartimento di Fisica e Astronomia,
  Universit\`a di Bologna, Viale Berti Pichat 6/2, I-40127 Bologna,
  Italy }
\affil{\altaffilmark{2} INAF - Osservatorio Astronomico di Bologna,
  via Ranzani 1, I-40127 Bologna, Italy }
\affil{\altaffilmark{3} Astrophysics Research Institute, Liverpool
  John Moores University, IC2, Liverpool Science Park, 146 Brownlow
  Hill, Liverpool L3 5RF, UK }
\affil{\altaffilmark{4} Max-Planck-Institute f$\ddot{u}$r
  Radioastronomie, D-53121 Bonn, Germany}

\begin{abstract}
We used ultra-deep UV observations obtained with the Hubble Space
Telescope to search for optical companions to binary millisecond
pulsars (MSPs) in the globular cluster 47 Tucanae. We identified four
new counterparts (to MSPs 47TucQ, 47TucS, 47TucT and 47TucY) and
confirmed those already known (to MSPs 47TucU and 47TucW). In the
color magnitude diagram, the detected companions are located in a
region between the main sequence and the CO white dwarf cooling
sequences, consistent with the cooling tracks of He white dwarfs of
mass between $\rm 0.15 \ \Msun$ and $\rm 0.20 \ \Msun$.  For each
identified companion, mass, cooling age, temperature and pulsar mass
(as a function of the inclination angle) have been derived and
discussed. For 47TucU we also found that the past accretion history
likely proceeded in a sub-Eddington rate. The companion to the redback
47TucW is confirmed to be a non degenerate star, with properties
particularly similar to those observed for black widow systems.
  Two stars have been identified within the 2$\sigma$ astrometric
  uncertainty from the radio positions of 47TucH and 47TucI, but the
  available data prevent us from firmly assessing whether they are the
  true companions of these two MSPs.
\end{abstract}


\section{INTRODUCTION}\label{intro}

Millisecond pulsars (MSPs) are rapidly spinning neutron stars (NSs)
formed in a binary system where a slowly rotating NS is spun up
through mass accretion from an evolving companion star.  The recycling process is usually observed in the low
mass X-ray binary systems, which are commonly considered as the MSP
progenitors  \citep{alpar82, batta91,papitto13}. Indeed, when the mass accretion rate decreases and the NS
is sufficiently recycled, the rotation-powered emission is
switched on and the system is observable as a MSP in the radio
band. These processes usually lead to a deep transformation of the
companion star which can transit through a highly perturbed
evolutionary phase \citep[possibly like MSP-A in NGC
  6397;][]{ferraro01a}, before reaching the final stage of a (possibly
He) white dwarf \citep[WD; e.g. MSP-A in NGC 6752;][]{fer03_msp}. Each
step of this evolution corresponds to objects characterized by
different properties, that, in the optical bands, are imprinted in the
observable features of the companion star.

According to the canonical scenario, the majority of binary MSPs have
low-mass He WD companions \citep[see, e.g.,][]{vankerkwijk05}. However, recent pulsar (PSR) searches have
considerably increased the number of non-canonical systems, especially
the so-called ``black widows" and ``redbacks": ultra-compact binary
systems ($\rm P_{ORB} \lesssim1 $ day) where the presence of radio
eclipses suggests the presence of ionized material ablated from a
bloated companion star because of the energy injected by the PSR
\citep{ruderman89, ray12, roberts13}. Redback companion stars usually have masses of
$\rm 0.1-0.5 \ \Msun$, while black widow companions are much less
massive ($\rm M<0.1 \ \Msun$). Such a small value could be due to
vaporization from the strong MSP radiation and relativistic wind. The
physical mechanisms bringing to one or the other system are still
debated. Simulations by \citet{chen13} show that redbacks and black
widows are the outcome of different evolutionary paths, where the PSR
irradiation efficiency is the discriminant factor. At odds with these
results, the simulations by \citet{benvenuto14} show that the
evolution of redbacks is bifurcated, with some of them evolving into
black widows, and the others producing canonical He WD
systems. Possibly, the progressive evaporation of the black widow
companions could lead to the total disruption of the star and then to
the formation of isolated MSPs. Interestingly, in recent years several connections between low mass X-ray binaries and redbacks have been found, especially with the discovery of systems transitioning from one state to the other \citep[see][]{archibald09, papitto13, bassa14}.

Although the Galaxy is $\sim100$ times more massive than the entire
Galactic globular cluster (GC) system, about 40\% of the known MSP
population is found in GCs. Such an over-abundance is indicative of a
strongly enhanced dynamical activity in these dense stellar systems,
which promotes the formation of a conspicuous number of exotic
objects, such as blue straggler stars, X-ray binaries, cataclysmic
variables and MSPs \citep{bailyn92, cool95, ferraro95, ferraro01b,
  ferraro15a, grindlay02, pooley03, ransom05}, which can be used to probe the
complex interplay between dynamics and stellar evolution
\citep[e.g.][]{goodman89, hut92, phinney92, possenti03, fer03_dyna,
  ferraro09, ferraro12,ferraro15b, verbunt14}. In this respect, the study of optical companions to binary MSPs in GCs is of outmost importance, since it opens the possibility to get insights on the impact of dynamical interactions (which are particularly frequent in dense environments) on MSP and stellar evolution, e.g., favoring binary formation (through tidal captures), binary shrinking (through fly-by) and consequent mass transfer activity, as well as exchange interactions able to substitute the original companion that recycled the pulsar, with a new, more or less perturbed, star \citep[see e.g.][]{rasio00, king03,fer03_mass,sabbi03a,sabbi03b,mucciarelli13,benaquista13}.  Moreover, in
the case of WD companions, it is possible to estimate the masses and
cooling ages of the systems by the direct comparison of their
properties with stellar evolutionary models \citep[see e.g.][]{fer03_msp,pallanca13a}, while accurate mass measurements require spectroscopical techniques \citep[e.g.][]{vankerkwijk96, bassa06, antoniadis12,antoniadis13}. In the case of MSPs in GCs, these constraints benefit from the known GC distance and optical extinction, thus reducing the uncertainties on the estimated quantities with respect to the case of MSPs in the Galactic field. The derived companion
masses can be combined with radio timing parameters to estimate the
PSR masses, thus allowing general relativity test
\citep[e.g.][]{freire12} and fundamental physics studies, as the
determination of the equation of state of ultra dense matter
\citep{lattimer01}. On the other hand, the derived cooling ages are a
more appropriate measurement of the system age with respect to the
characteristic PSR ages derived from radio timing
\citep{tauris12a,tauris12b}. The correct determination of MSP ages is
an important tool to study the spin evolution and to constrain the
physics of the recycling phases \citep[see e.g.][and references
  therein]{vankerkwijk05}.

Despite their importance, the identification of MSP optical companions
is challenging in crowded stellar systems like GCs. Only ten
companions have been discovered so far in GCs. Three companions are He
WDs \citep[see][]{edmonds01, fer03_msp, sigurdsson03,bassa03,bassa04}, as expected
from the canonical formation scenario, five are redbacks companions
\citep[see][]{ferraro01a, edmonds02, cocozza08, pallanca10,
  pallanca13b} and two are black widow companions
\citep[][]{pallanca14a,cadelano15}.

The GC 47 Tucanae, located at a distance of about $4.5$ kpc from the
Sun, hosts the largest population of MSPs after Terzan 5
\citep{freire03,ransom05}. Indeed, 23 radio MSPs have been discovered
so far, 14 of which are located in binary systems\footnote{Please
  visit \url{http://www.naic.edu/~pfreire/GCpsr.html}, for a complete
  list of the main radio timing properties of MSPs in GCs.}. Here we
report the identification and the properties of four new MSP
companions in 47 Tucanae and we present the follow-up study of two
previously known companions. In Table~\ref{tab1} we report the main
radio timing properties of the analyzed objects, which are useful in
the following discussions. All the identified companions are
likely canonical MSPs, except one which is a redback system.

In Section 2 we present the used photometric dataset and the
identification of the MSP optical counterparts. In Section 3 we
discuss in detail the properties of each companion. Finally, in
Section 4, we summarize our results.

\section{OPTICAL PHOTOMETRY OF THE STAR CLUSTER}
\label{identification}
\subsection{Observations and data analysis}
\label{Sec:dataan}
In the present study the identification of the MSP companions has been
performed through an ultra-deep, high resolution, photometric dataset
acquired under GO 12950 (P.I: Heinke) with the UVIS camera of the Wide
Field Camera 3 (WFC3) mounted on the Hubble Space Telescope (HST). The
dataset consists of 8 images in the F390W filter, with exposure times
of $567-590$~s, and 24 images in the LP F300X filter, with exposure
times of $604-609$~s.
 
The standard photometric analysis \citep[see][]{dalessandro08a,
  dalessandro08b} has been performed on the ``flt'' images, which are
corrected for flat field, bias and dark counts.  These images have
been further corrected for ``Pixel-Area-Map''\footnote{For more
  details see the WFC3 Data Handbook.} with standard IRAF
procedures. By using the {\textrm DAOPHOT II} packages
\citep{stetson87}, we performed an accurate photometric analysis of
each image. First of all, we modeled a spatially varying Point Spread Function (PSF)
by using a sample of $\sim200$ bright but not saturated stars. The
model has been chosen on the basis of a $\chi^{2}$ test and, in every
image, the best fit is provided by a Moffat function
\citep{moffat69}. Then we performed a source detection analysis,
setting a $3\sigma$ detection limit, where $\sigma$ is the standard
deviation of the measured background.  Once a list of stars was
obtained, we performed a PSF-fitting in each image using the {\textrm
  ALLSTAR} routine. In the resulting catalog we included only objects
present at least in half the images for each filter. Then, this
catalog has been further processed with the {\textrm ALLFRAME}
routine. For each star, we homogenized the magnitudes estimated in
different images, and their weighted mean and standard deviation have
been finally adopted as the star mean magnitude and its related
photometric error \citep[see][]{ferraro91, ferraro92}. However, in
order to perform variability studies, for each source we also kept the
homogenized magnitude measured in each frame in both filters. Then,
instrumental magnitudes have been calibrated to the VEGAMAG system by
using the zero points quoted in the WFC3 Data Handbook and by
performing aperture corrections.

\subsection{Astrometry}

Since the WFC3 images suffer from geometric distortions, we corrected
the instrumental positions (x,y) following \citet{bellini11}. In order
to transform the instrumental positions into the absolute astrometric
system ($\alpha, \delta$), we used, first of all, the wide field
  catalog presented in \citet{ferraro04}. Its astrometric solution
has been improved by cross-correlation\footnote{We used CataXcorr, a
  code which is specifically developed to perform accurate astrometric
  solutions. It has been developed by P. Montegriffo at INAF-
  Osservatorio Astronomico di Bologna. This package is available at
  http://davide2.bo.astro.it/$\sim$paolo/Main/CataPack.html, and has
  been successfully used in a large number of papers by our group in
  the past 10 years.} with the UCAC4 astrometric standard catalog (\citealp{zacharias13}; $\sim4600$ stars have been found in common
  between the two datasets). The latter is based on the
International Celestial Reference System, thus allowing a more
appropriate comparison with the MSP positions derived from timing
using solar system ephemerides (which are referenced to the same
system). The newly-astrometrized wide field catalog has then
been used as a secondary reference frame to astrometrize the WFC3
  data set, by means of $\sim22000$ stars in common. The resulting
  $1\sigma$ astrometric uncertainty is $0.10\arcsec$ and $0.11\arcsec$
  in $\alpha$ and $\delta$, respectively. Thus the final total
  astrometric uncertainty is $\sim0.15\arcsec$. Unfortunately, there
are only few stars in common between the WFC3 and the UCAC4 catalogs,
since the latter does not cover the cluster central regions. This
prevented a direct cross-correlation between the two catalogs and thus
we could not take into account the stellar proper motions between the
two observation epochs, which would have reduced the astrometric
uncertainty.

\subsection{Identification of the MSP companions}

\begin{figure*}
\begin{center}
\leavevmode
\includegraphics[width=4.1cm]{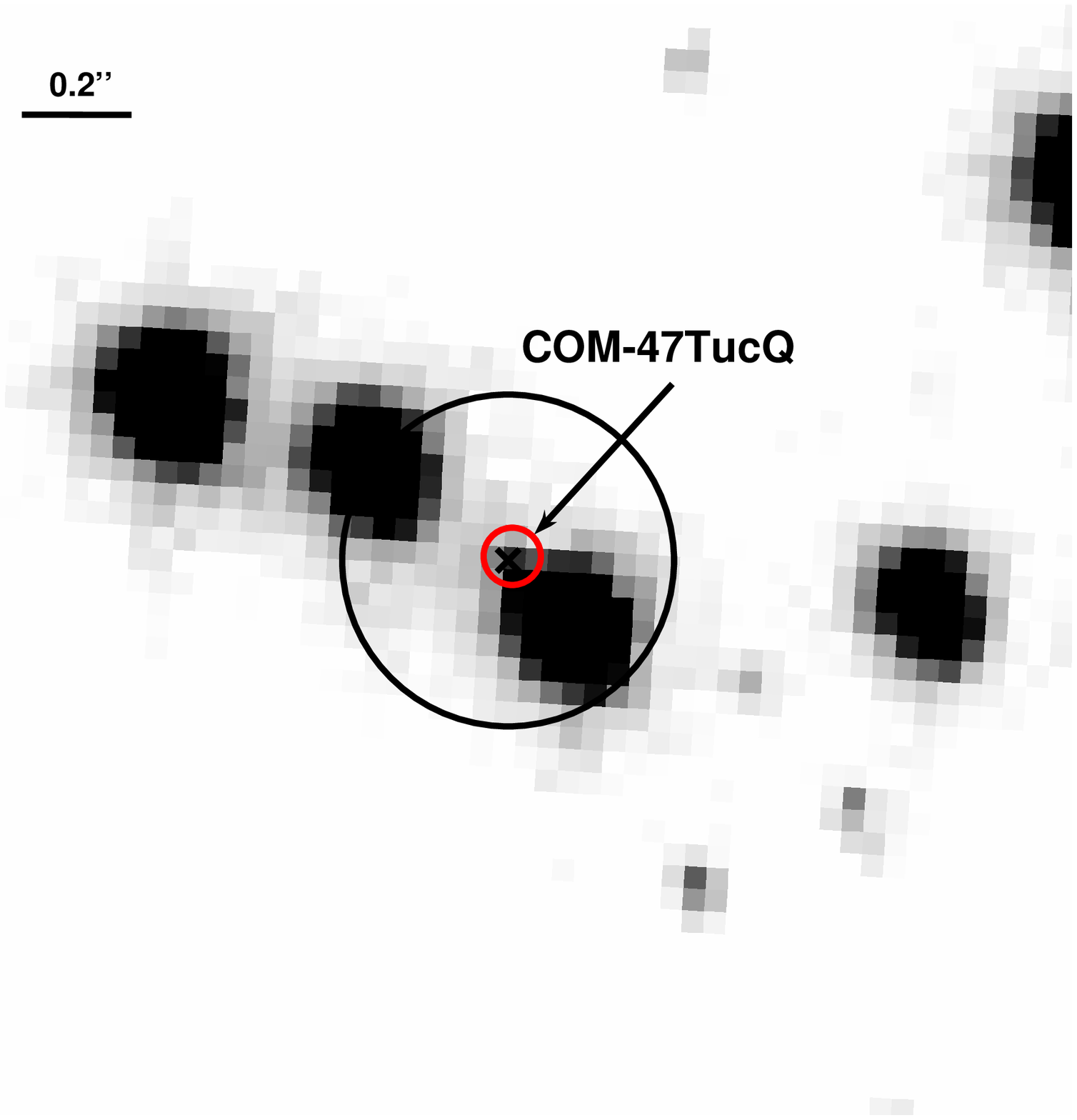}
\includegraphics[width=4.1cm]{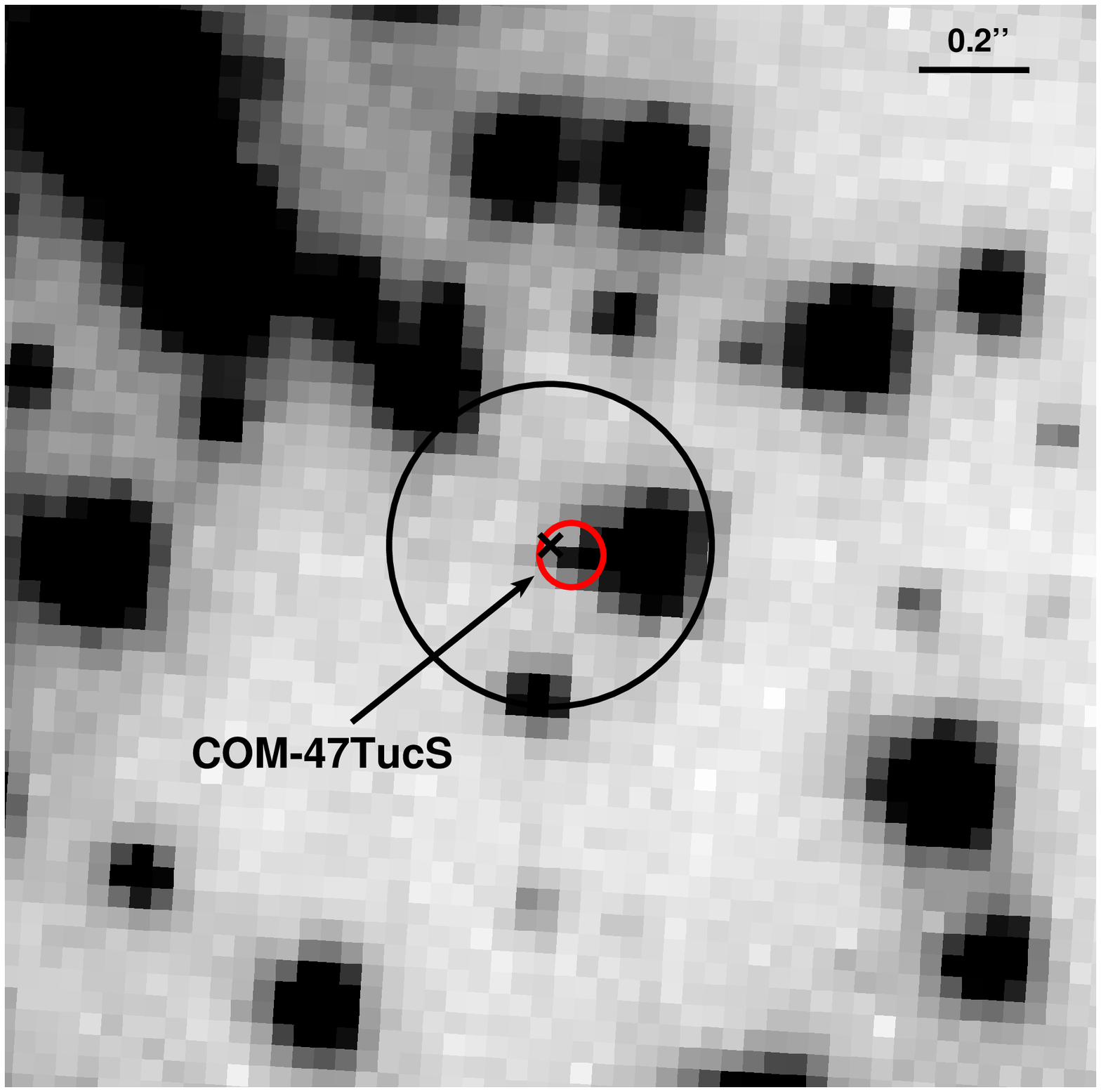}
\includegraphics[width=4.1cm]{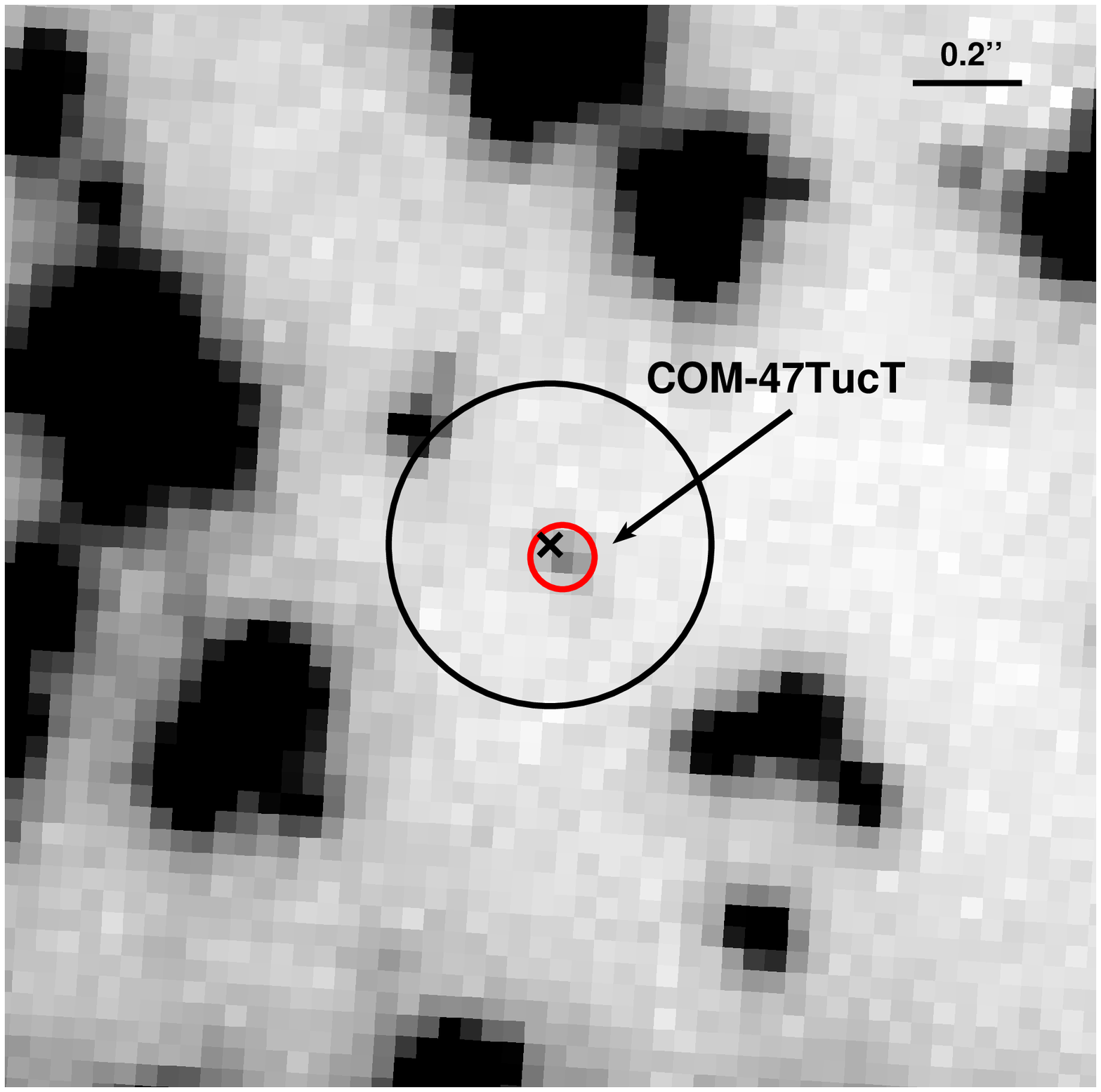}
\includegraphics[width=4.1cm]{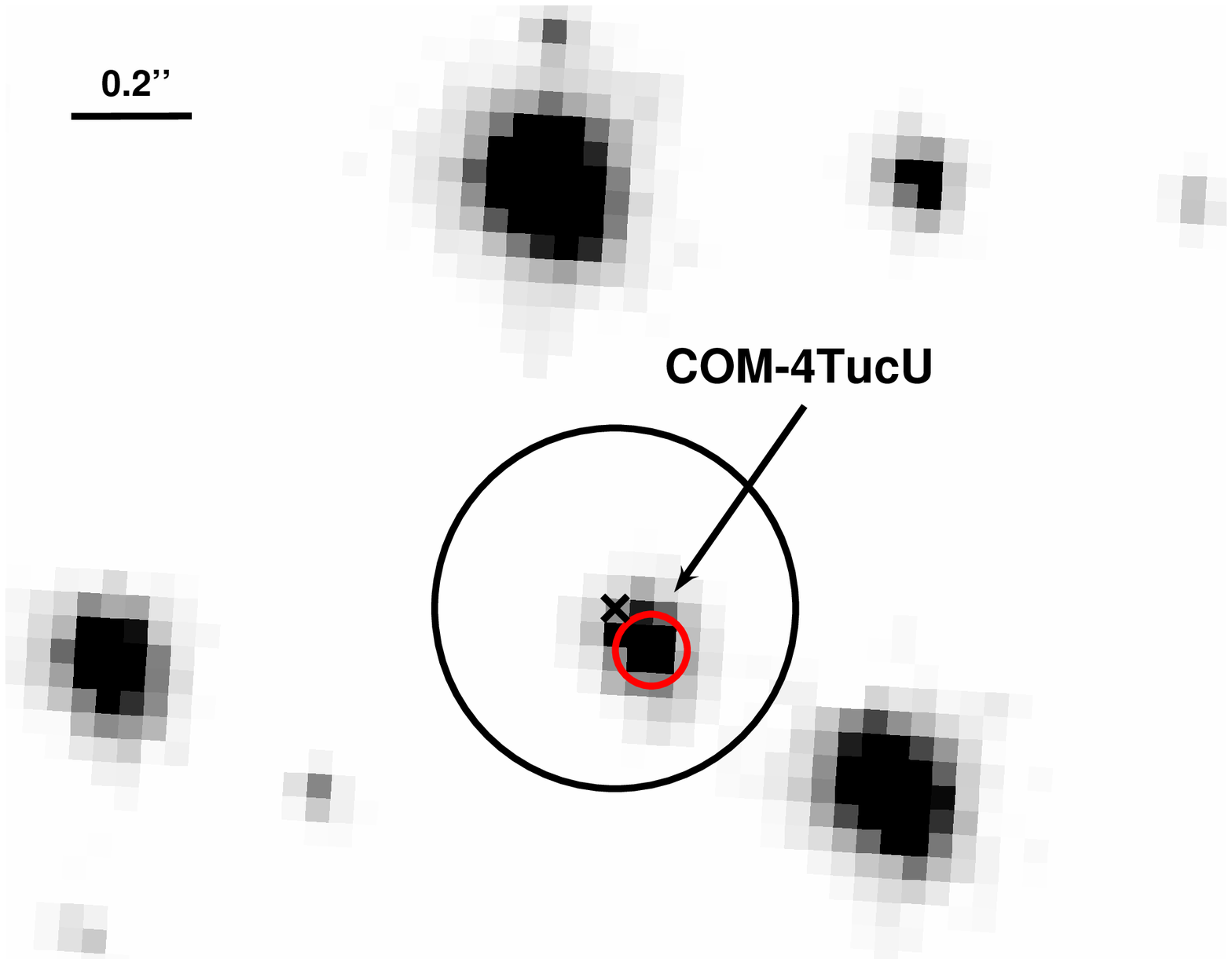}
\includegraphics[width=4.1cm]{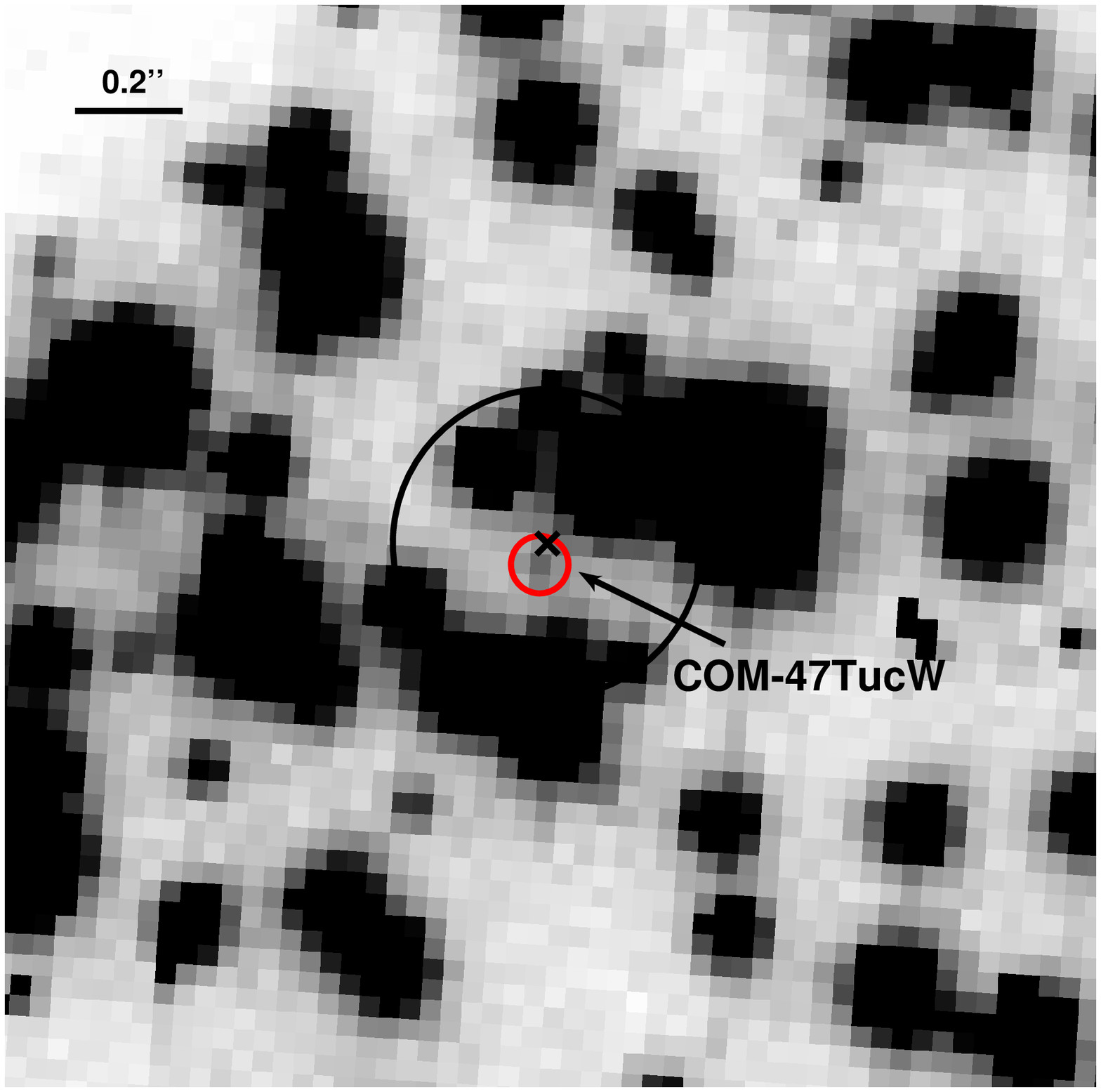}
\includegraphics[width=4.1cm]{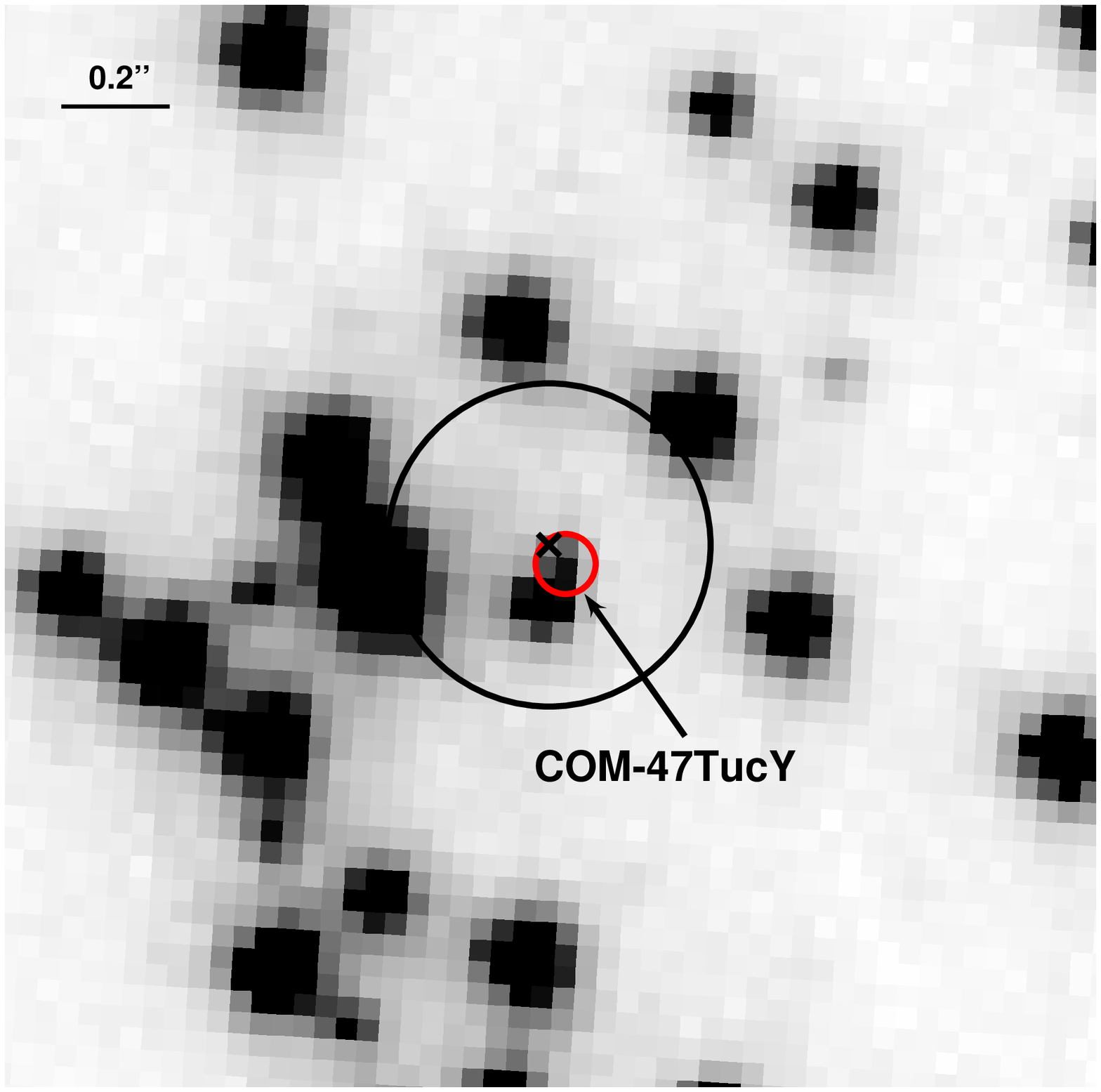}
\includegraphics[width=4.1cm]{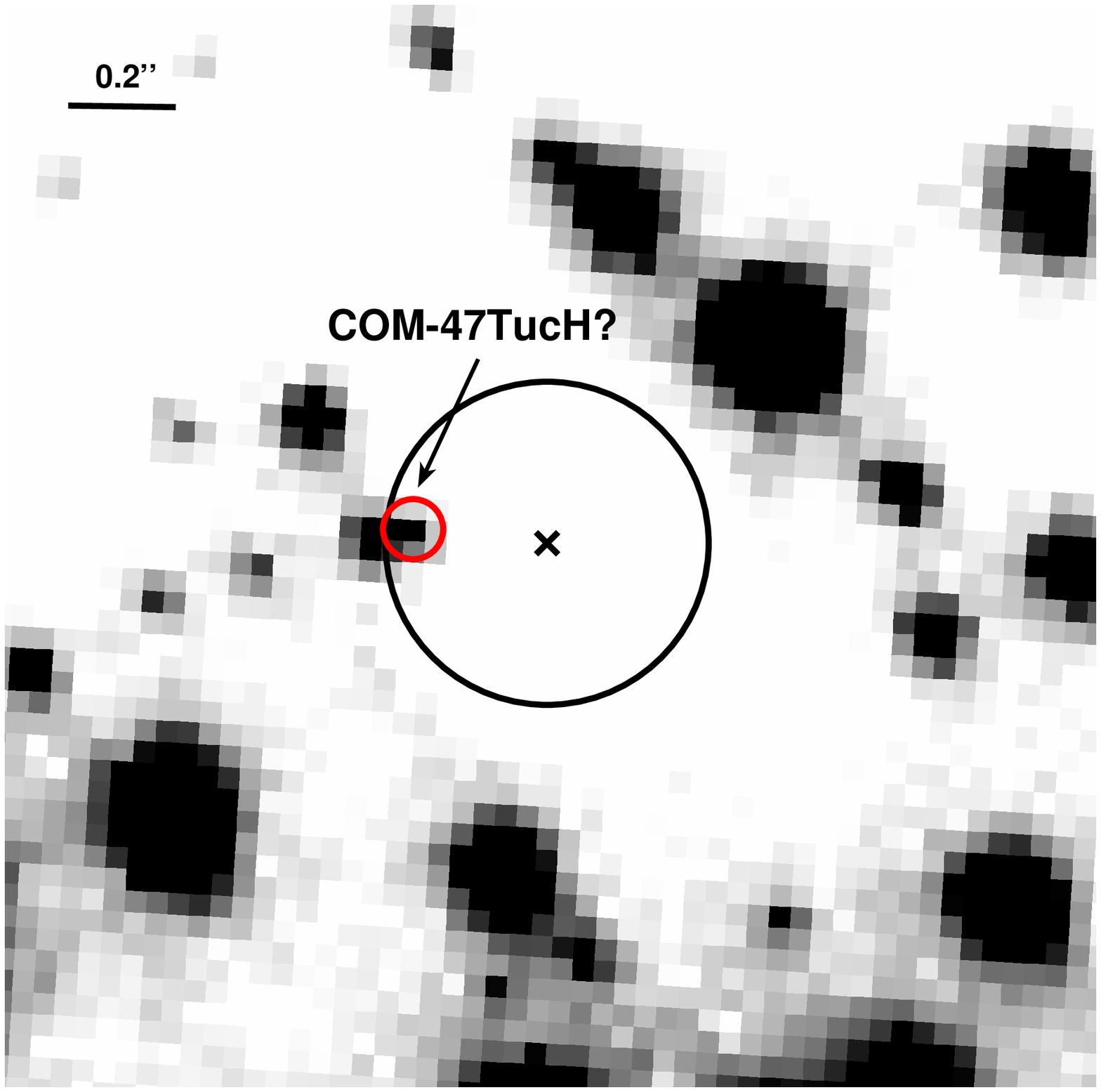}
\includegraphics[width=4.1cm]{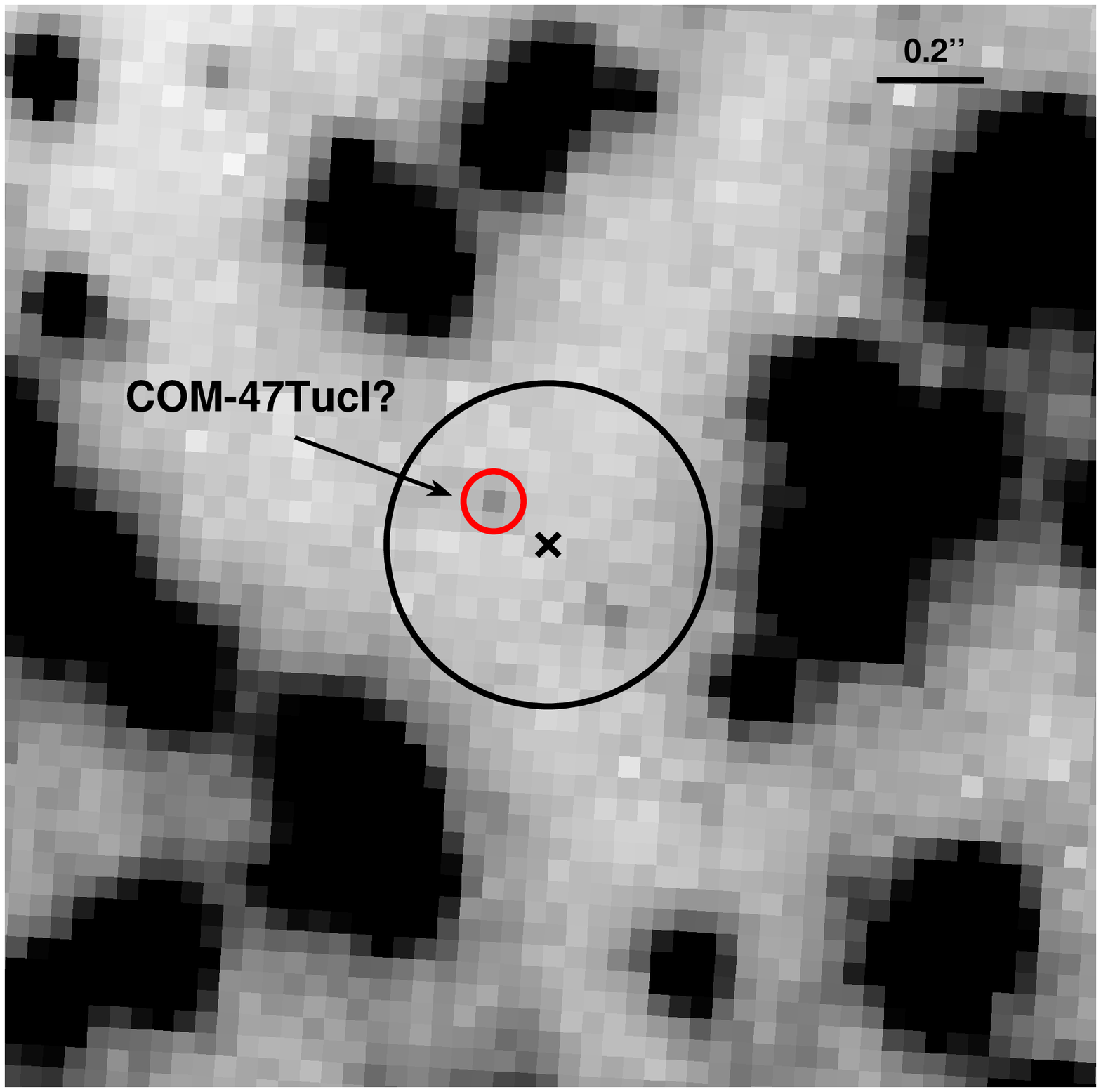}
\caption{HST images of the $2\arcsec \times 2\arcsec$ region around
  the nominal position of the seven MSPs analyzed in this work. North
  is up and east is left. All the charts are obtained from a
  combinations of the available F300X images, with the exception of
  that of 47TucW that is from an image where the companion star is
  at its maximum luminosity. The black circles are centered on the
  radio PSR nominal position in the optical astrometric system and
  their radii are equal to our $2\sigma$ astrometric uncertainty
  ($0.30\arcsec$). The red circles mark the identified MSP companions.}
\label{fig1}
\end{center}
\end{figure*}

First of all, in order to search for the companions to the MSPs in 47
Tucanae, we checked the precision of our astrometric solution
re-identifying the two companion stars already known in the cluster
\citep[see][]{edmonds01,edmonds02}. To this aim, we performed a
detailed analysis of all the detectable objects within a $5\arcsec
\times 5\arcsec$ wide region centered on the nominal position of each
MSP. The companion to 47TucU (COM-47TucU; hereafter all the companions
will be named as COM-47Tuc followed by the letter of the respective
MSP) and 47TucW have been re-identified in stellar sources located at
$0.06\arcsec$ from the MSP nominal positions. Both the identifications
turn out to be largely within our astrometric uncertainty, thus
confirming the accuracy of the adopted astrometric solution. The
finding charts of these two reference objects are shown in Figure
\ref{fig1}.

Following the same procedure, we searched for the companions to all
the other MSPs with a known position \citep[][Freire et al. 2015, in
  preparation]{freire03}.  Stars located within the $2\sigma$
  uncertainty from the pulsar position have been considered as
possibile counterparts. Four companions (to 47TucQ, 47TucS, 47TucT and
47TucY) have been identified on the basis of their positional
coincidence (all of them are located at a distance $\leq0.06\arcsec$
from the nominal radio position) and of their position in the color
magnitude diagram (CMD). Two faint stars have been detected also
  within the $2\sigma$ uncertainty circle from 47TucI and
  47TucH. However, their distances ($0.15\arcsec$ and $0.24\arcsec$
  respectively) from the pulsar radio positions are significantly
  larger than in all the other cases, thus casting doubts about these
  objects being the true optical counterparts (see more discussion in
  Section~\ref {comH}). The finding charts of all these objects are
shown in Figure~\ref{fig1} and their main photometric properties are
reported in Table~\ref{tab2}. Their location in the cluster CMD
is shown in Figure~\ref{cmd}, where only the stars with a sharpness
parameter\footnote{The sharpness parameter is a {\textrm DAOPHOT II}
  output that quantifies the stellar-like structure of each object
  fitted with the PSF model. See the User Manual for more details.}
$\rm |sh|\leq0.05$ are plotted. As can be seen, with the
  exception of the candidate companion to 47TucI, all the newly
identified counterparts are located in the region where He WDs are
expected, although the candidate companion to 47TucH could be
compatible also with the CO WD cooling sequence (see Section
  \ref{comH}). Since the radio timing properties suggest that these
systems are the product of the canonical recycling scenario, their
location along the He WD cooling sequences guarantees their connection
with the MSPs. Note in fact that the probability of a chance
coincidence with another He WD is extremely low ($\rm
P\approx0.1\%$)\footnote{The chance coincidence probability has been
  evaluated predicting the number He WD expected within a radius equal
  to the $2\sigma$ astrometric uncertainty. We derived the He WD
  density by direct counting of the all objects located among the
  cooling tracks (see Section~\ref{prop} and Figure~\ref{cmdwd}) and
  dividing this number by the size of the WFC3 field of view. Please
  note that even including all the stars of the catalog with sharpness
  $\rm |sh|>0.05$, the chance probability remains $\lesssim0.5\%$.},
since these objects can only be the product of the late stage of the
evolution of exotic objects like, for example, MSPs and cataclysmic
variables. The candidate companion to 47TucI is instead a main
  sequence-like object, and its properties will be briefly discussed
  in Section \ref{comH}.  As concerns the previously known
companions, COM-47TucU is also located along the He WD sequence, while
the redback COM-47TucW is located in an anomalous region between the
main sequence and the WD cooling sequence (see Section~\ref{w}).

With the exception of the COM-47TucW, no significant variability
related to the orbital period has been detected. For 47TucU, 47TucY
and 47TucW (see Section~\ref{w}) the observations sample a significant
fraction of the orbital period. Instead, for the other systems (with
orbital periods longer than 1 day) the coverage is too poor to allow
any appropriate variability analysis.  However, a strong magnitude
modulation, as the one observed for non degenerate companions
\citep[see e.g.][]{stappers99,edmonds02,reynolds07, pallanca10,
  romani11, pallanca14a,cadelano15}, is not expected and usually not
observed for degenerate objects, since the flux enhancement due to
re-heating of the companion star by the PSR emitted energy is
negligible.

\section{DISCUSSION}

\subsection{The physical properties of the He WD companions}
\label{prop}
In order to constrain the main properties of the He WD companions, we
have compared the position of each candidate in the CMD with a set of
He WD cooling tracks computed by \citet{althaus13}. These models span
a mass range from $\rm 0.15 \ \Msun$ to $\rm 0.43 \ \Msun$, spaced at
about $0.005 \ \Msun$ for masses between $\rm 0.15 \ \Msun$ to $\rm
0.19 \ \Msun$ and up to $\rm 0.07 \ \Msun$ for larger masses. We
transformed the theoretical luminosities and temperatures into the
absolute F300X and F390W magnitudes, by applying the bolometric
corrections kindly provided by P. Bergeron
\citep[see][]{holberg06,bergeron11}.  Then, the model absolute
magnitudes have been transformed into the apparent ones by using the
distance modulus $(m-M)_{0}=13.32\pm0.10$ \citep{ferraro99}\footnote{
  Many literature works reported on different values of 47 Tucanae
  distance modulus \citep[see, e.g.][and references
      therein]{woodley12}. However, all these possibile values have
  only a minimal influence on our derived companion properties
    (e.g. the derived companion masses would vary of less than
    $\sim7\%$ for all the companion stars)}. and the color excess
$E(B-V)=0.04\pm0.02$ \citep{ferraro99,zoccali01,salaris07} and
extinction coefficients $A_{F300X}/A_{V}=1.77309$,
$A_{F390W}/A_{V}=1.42879$
\citep[][]{cardelli89,odonnel94}. Figure~\ref{cmdwd} shows the zoomed
portion of the CMD in the WD region with a sample of cooling tracks
for different masses overplotted. As can be seen, the range in mass of
the models is large enough to properly sample the portion of the CMD
where all the companions are located. Therefore we used this set of
models to derive the combinations of parameters (mass, cooling age and
temperature) that simultaneously satisfy the observed photometric
magnitudes in both the filters, also taking into account the
uncertainties on the companion magnitudes, distance modulus and
reddening. The best values have been evaluated with a simple
$\chi^{2}$ statistic. In doing this, linear interpolations (for
different masses but equal ages) among the tracks have been performed
in order to have a tighter mass sampling. We assumed that each
companion is located at the distance of 47 Tucanae\footnote{Even
  though \citet{freire01a} measured distance offsets between the
  cluster MSPs, such differences are very small and can be neglected
  for our goals.} and it is affected by the same extinction\footnote{
  The effects of differential reddening are negligible for our goals
  \citep[see][]{milone12a,milone12b}.}. Figure \ref{bo1} shows, for
each system, the combination of cooling age (left panel), temperature
(central panel) and PSR mass (right panel) appropriate for the derived
value of the companion mass. In particular, in each plot the right
panel shows the results obtained for different values of the
inclination angle and interesting constrains on each system can be
drawn. For instance, by setting the inclination angle to $90^{\circ}$,
the maximum PSR mass allowed from the inferred companion mass can be
evaluated. Conversely, by assuming the minimum PSR mass equal to $\rm
1.17 \ \Msun$ \citep[the lowest mass ever measured for a
  NS;][]{janssen08}, a conservative lower limit to the inclination
angle can be derived. All these results are also summarized in
Table~\ref{tab3} where the quoted uncertainties are the range of
possibile values allowed by the comparison with the theoretical
tracks\footnote{The reader should be aware that the WD parameters
  should not be assumed at face value as perfectly correct but as
  estimations, since they are model dependent and could also suffer
  from some hardly quantifiable uncertainty linked to the bolometric
  corrections.}. Note that we are not analyzing here the cases of
  47TucH and 47TucI, which we will discuss in Section~\ref{comH}.

As can be seen, all the companions have masses between $\rm \sim0.15
\ \Msun$ and~$\rm \sim0.2 \ \Msun$. The derived ranges of ages are in
agreement with the lower limits to the PSR characteristic ages
reported in Table~\ref{tab1}. The only exception is COM-47TucU, which
is discussed below. In principle the mass of COM-47TucQ could be
smaller than our best-fit value ($0.15 M_\odot$), since not
theoretical tracks for masses below this value are available. However,
already a $0.15 \ \Msun$ companion would imply an extremely low value
of the PSR mass ($\rm \lesssim 1 \ \Msun$). This puzzling result could
be partially explained with the difficulty of accurately determine the
color of the optical counterpart, because of the presence of a very
close bright object (see Figure~\ref{fig1}).

Our results rule out a massive NS in the case of 47TucQ and
47TucT, while the possibility of a $\rm \sim2 \ \Msun$ NS remains
opened in the cases of 47TucS, 47TucU and 47TucY. However,
Figure~\ref{bo1} shows that the PSR mass can be significantly reduced
by assuming an intermediate-low inclination angle of the orbital
plane. In any case these systems are worthy of future, especially spectroscopical,
investigations. 

\begin{figure*}
\begin{center}
\leavevmode
\includegraphics[width=12cm]{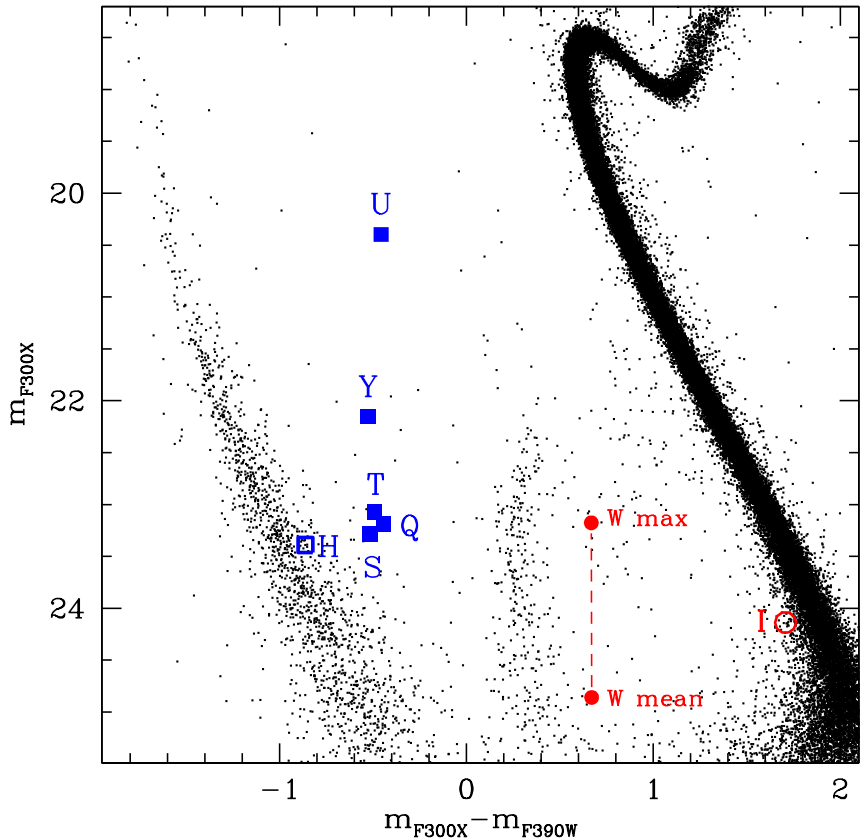}
  \caption{UV CMD of the GC 47 Tucanae. Only stars with sharpness
    parameter $\rm |sh|\leq0.05$ are plotted. The blue solid squares
    mark the companions to the canonical MSPs. The possible
    counterparts to 47TucH and 47TucI are plotted as an open square and circle respectively. Since
    COM-47TucW is a strongly variable object, we report its position
    at the maximum and mean luminosities, as derived by the best-fit
    models (see Section \ref{w} and Figure~\ref{curva}).}
  \label{cmd}
\end{center}
\end{figure*}

\begin{figure*}
\begin{center}
\leavevmode
\includegraphics[width=12cm]{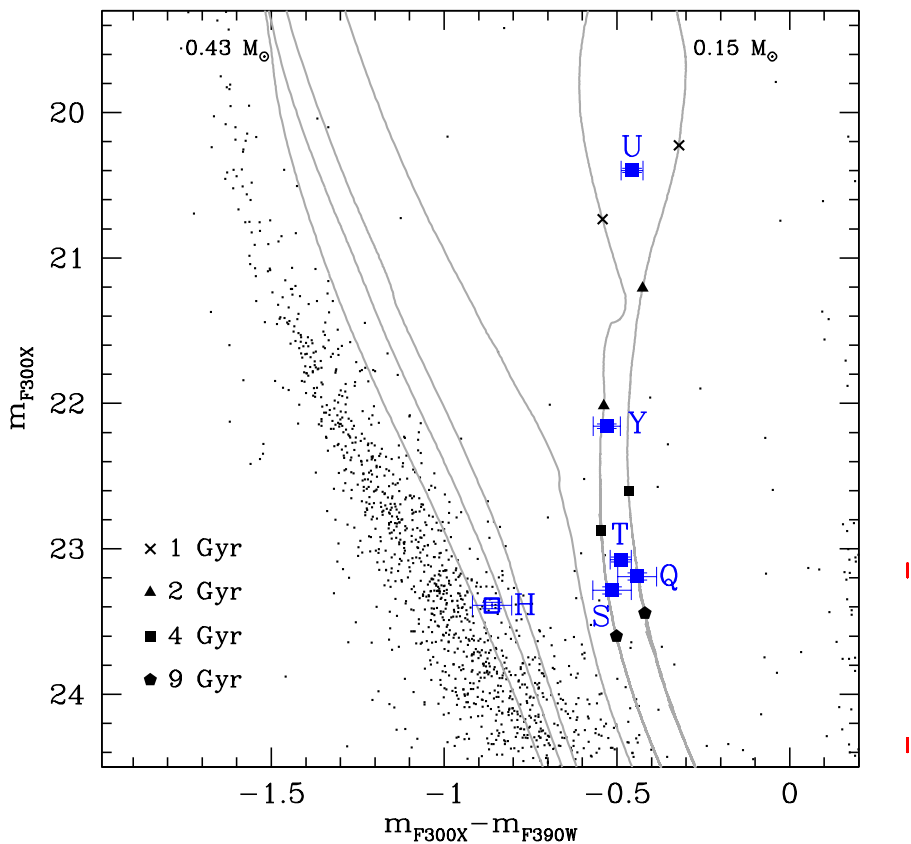}
  \caption{Same as in Figure~\ref{cmd}, but zoomed into the WD
    region. The continuous curves are reference He WD cooling tracks
    for stars of $\rm 0.15 \ \Msun$, $\rm 0.17 \ \Msun$, $\rm 0.20
    \ \Msun$, $\rm 0.32 \ \Msun$, $\rm 0.36 \ \Msun$ and $\rm 0.43
    \ \Msun$ (from right lo left). For the two rightmost tracks, points at 1,2,4 and 9 Gyrs have been marked with different symbols. The photometric errors of the
    companion stars are also drawn. }
  \label{cmdwd}
\end{center}
\end{figure*}

\begin{figure*}
\begin{center}
\leavevmode
\includegraphics[width=7cm]{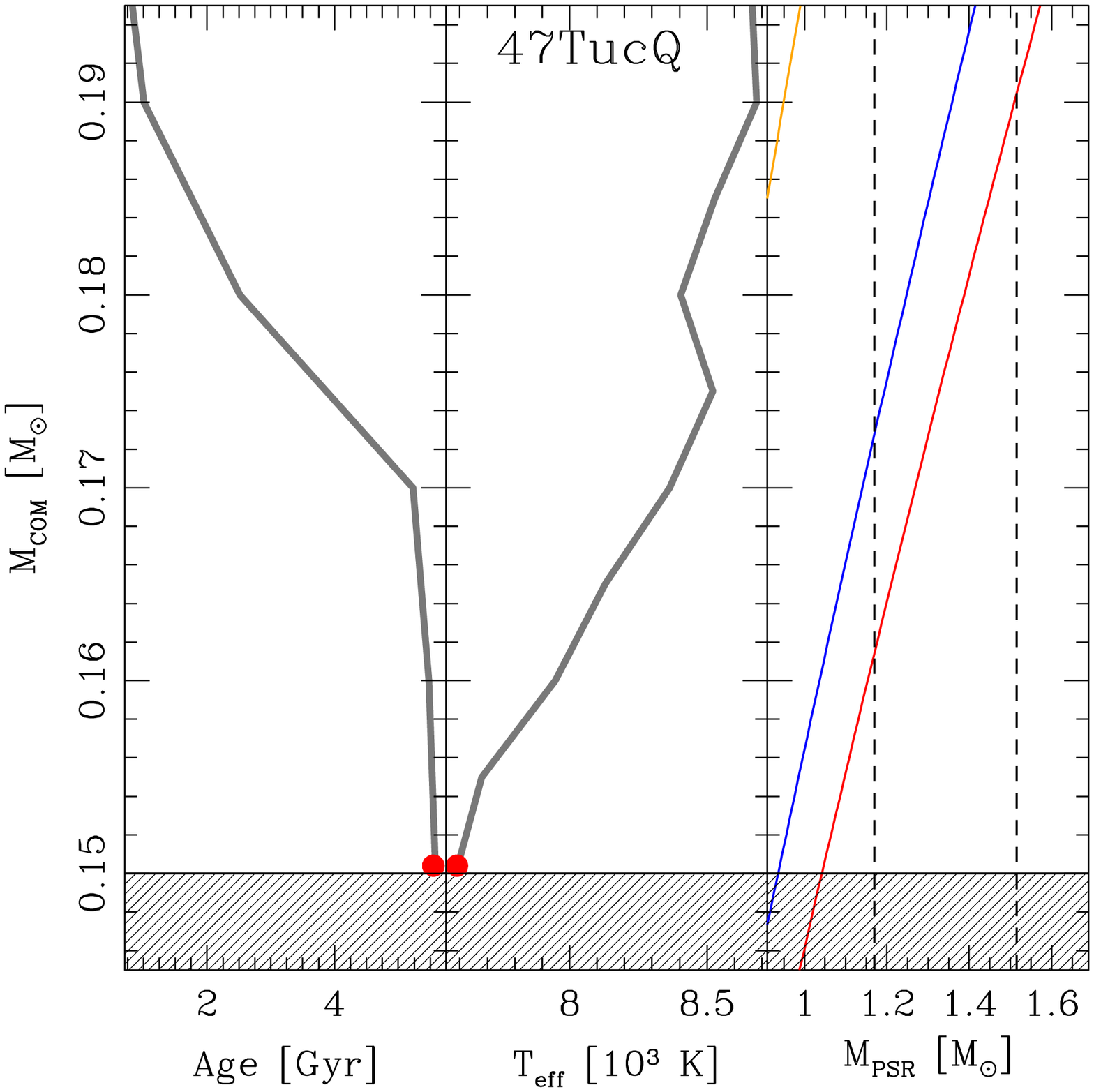}
\includegraphics[width=7cm]{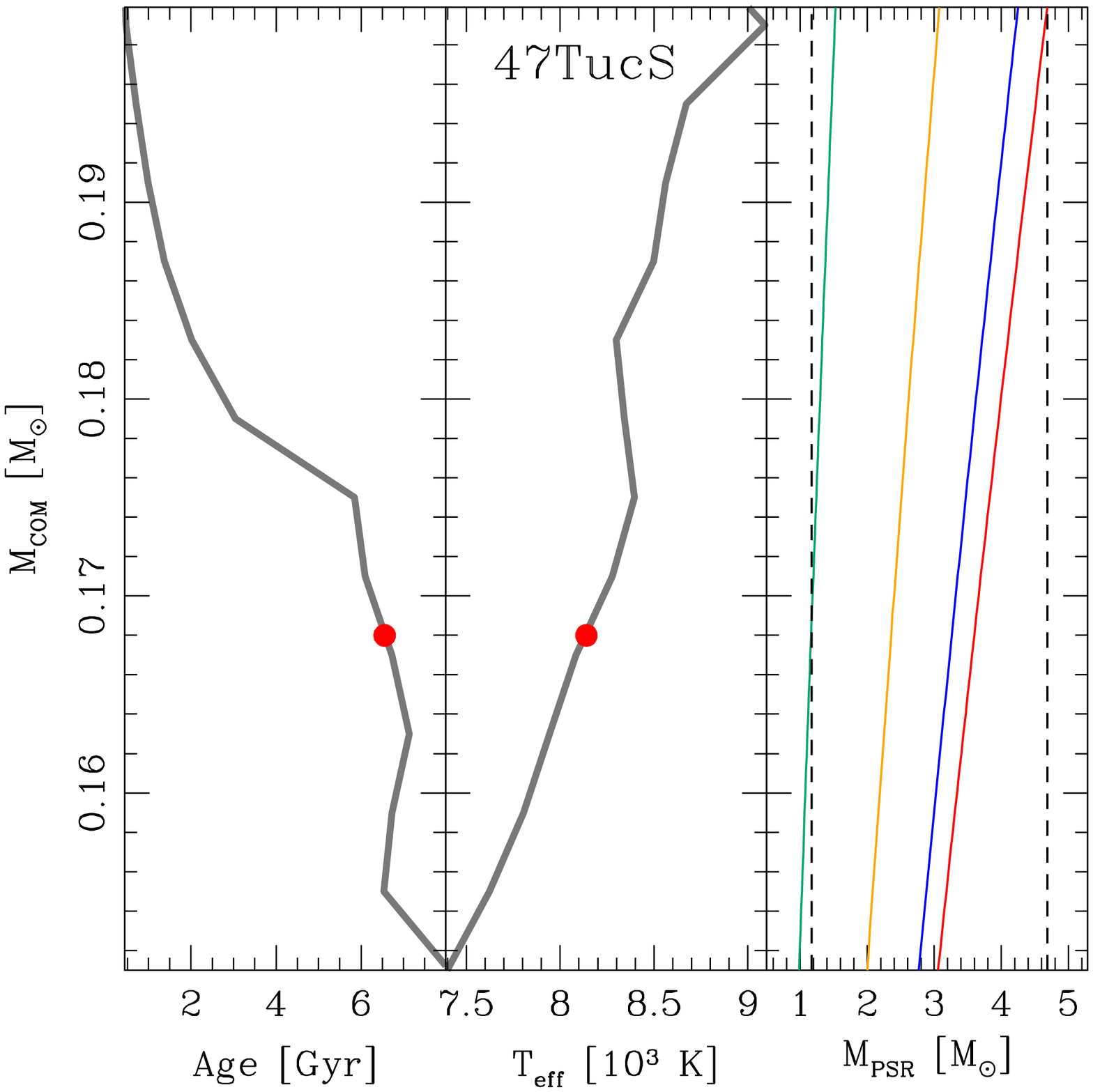}
\includegraphics[width=7cm]{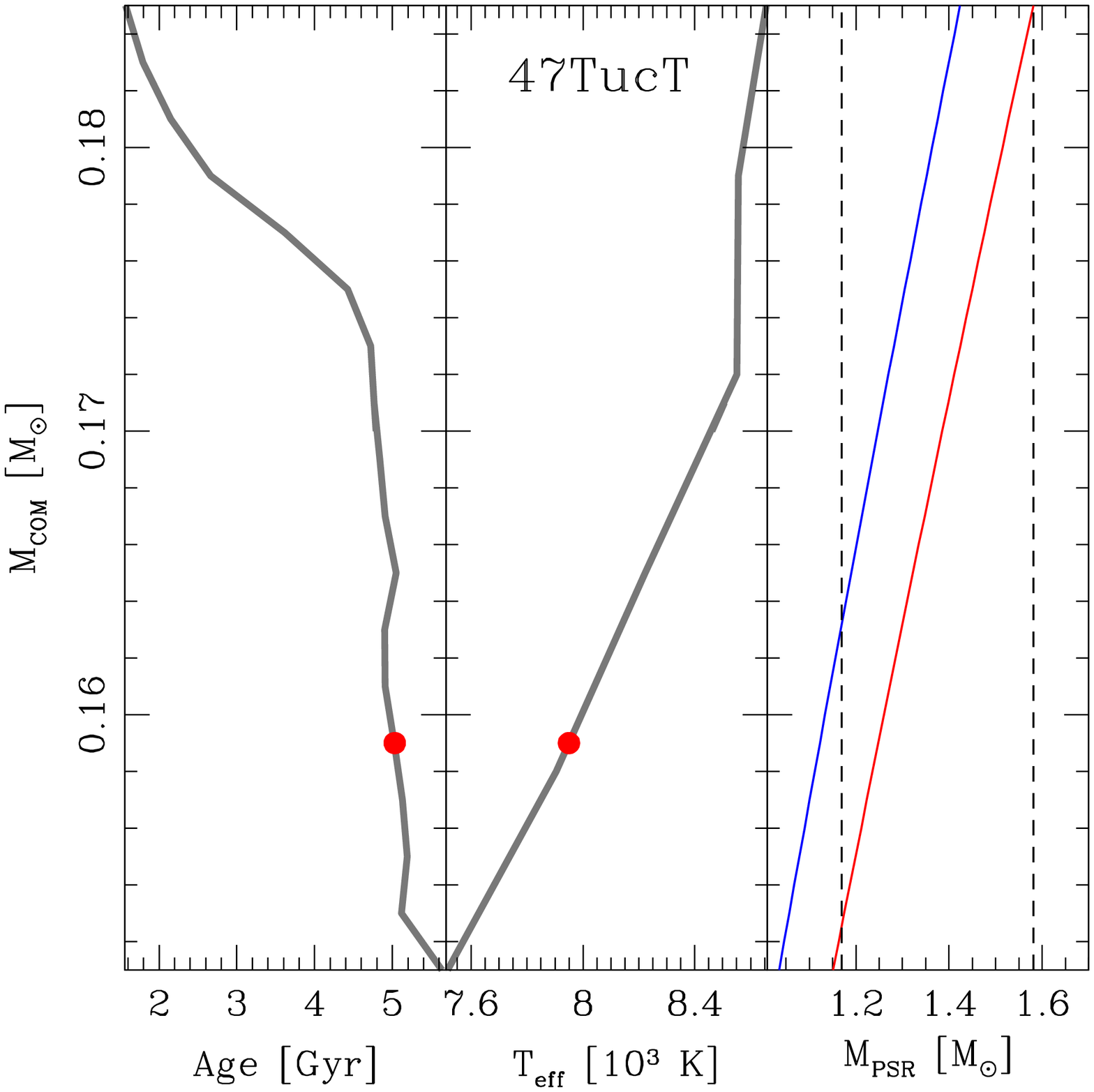}
\includegraphics[width=7cm]{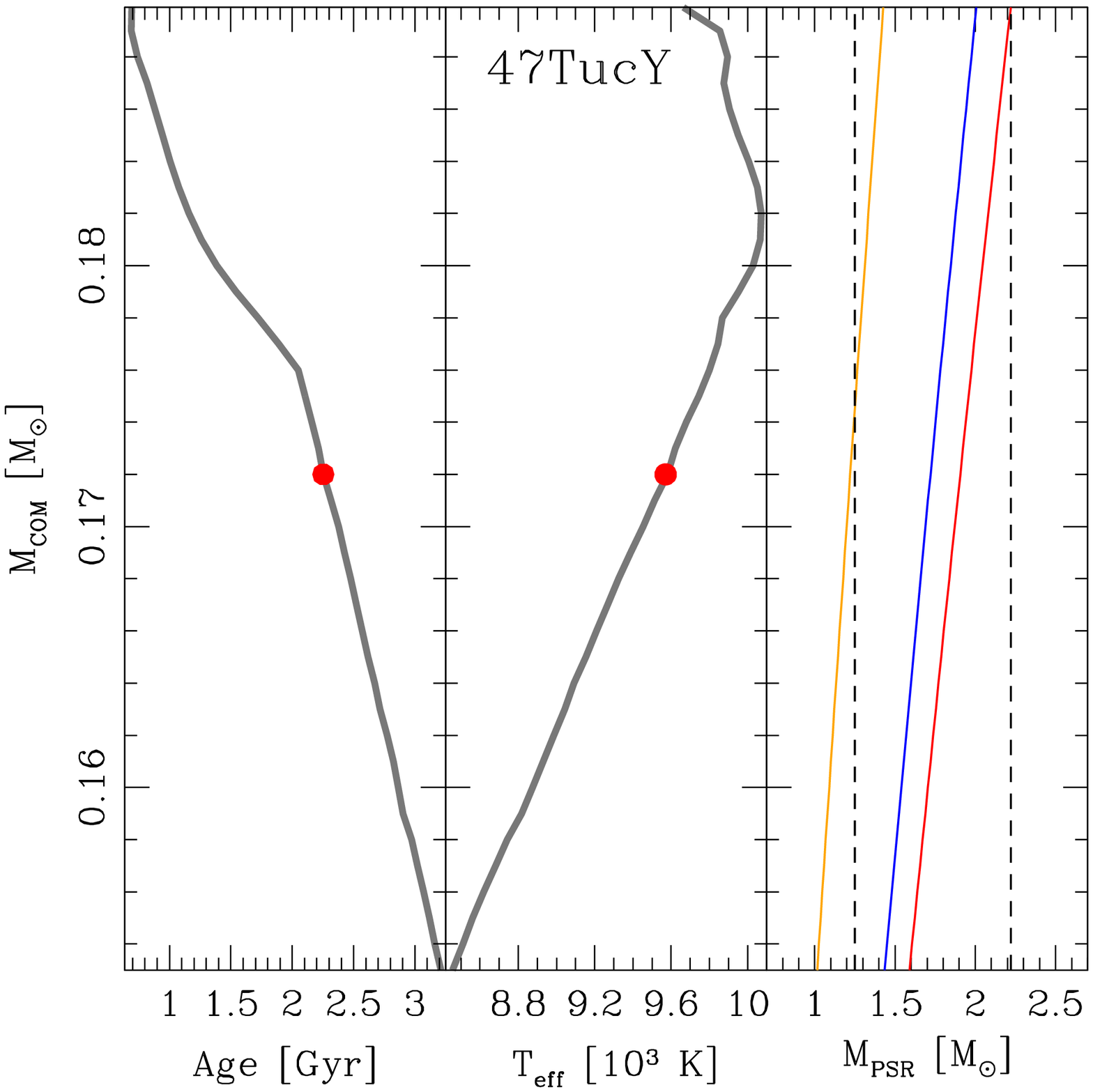}
  \caption{Physical properties of 47TucQ, 47TucS, 47TucT and
    47TucY (see labels), as derived from the comparison between the
    photometric characteristics of each companion and the WD cooling
    track models. In each plot, the gray lines drawn in the left and
    central panels show the allowed combinations between companion
    mass and cooling age or temperature (see text). The red dots
    correspond to the most probable values. In the case of COM-47TucQ
    the shaded areas mark the region ($\rm M_{COM}<0.15 M_\odot$) not
    sampled by the theoretical cooling tracks. In the rightmost panel
    of each plot, the solid curves represent the combination of values
    allowed by the PSR mass function for different inclination angles
    ($i=90^{\circ}$ in red, $i=70^{\circ}$ in blue, $i=50^{\circ}$ in
    orange, and $i=30^{\circ}$ in green). The blue dashed lines
    correspond to the assumed minimum NS mass \citep[$\sim 1.17
      \Msun$;][]{janssen08} and the largest NS mass value obtained for
    $i=90^{\circ}$.}
\end{center}
\label{bo1}
\end{figure*}

We also compared our results with the theoretical predictions on the
behavior of the orbital period as a function of the companion mass
discussed in \citet{tauris99}. Such a model has been already
empirically verified by \citet{corongiu12} and \citet{bassa06}.  As shown in
Figure~\ref{mp}, where we also added the two He WD companions
identified in NGC 6752 and M4 \citep[see][]{fer03_msp,sigurdsson03},
our results are in reasonable agreement with the model. The analytical
prediction seems to slightly overestimate the companion mass or to
underestimate the system orbital period. However, this model is valid
for binary systems with $\rm 0.18\ \Msun<M_{WD}<0.45\ \Msun$, thus
only marginally representative of our sample, where most of the
companions appear to be less massive than $\rm 0.18 \ \Msun$. More
updated models (from \citealt{istrate14}; gray points in
Figure~\ref{mp}) are in better agreement, although they are the
results of simulations of donor stars with metallicity $\rm Z=0.02$,
larger than that of 47 Tucanae {\citep[i.e. $\rm Z=0.008$,][]{lapenna15}.

The brightness of COM-47TucU allowed us to put tighter constraints to
the system parameters with respect to the other objects. Both its mass
and temperature are in excellent agreement with those reported in
\citet{edmonds01}, while our derived age ($\approx 0.9$ Gyr) turns out
to be 0.3 Gyr larger than their estimate. Such a discrepancy could be
due to the different theoretical models used. However, as already
noticed by \citet{edmonds01}, the cooling age is significantly lower
than the characteristic age of 2.5 Gyrs\footnote{This value is based
  on the estimate of the PSR spin-down rate from the orbital period
  derivative, which is precise enough for this system (Freire et al.,
  in preparation).}. This discrepancy should not alarm, since the PSR
characteristic ages are based on many assumptions and large deviations
from the companion cooling ages are commonly observed \citep[see
  e.g.][]{handbook,tauris12a,tauris12b}.  Using the WD age together
with the intrinsic spin period derivative ($\rm
\dot{P}=2.7\pm0.5\times10^{-20}$; Freire et al., in preparation) and
the actual spin period ($\rm P\approx4.343$ ms), we evaluated a MSP
birth spin period (the so-called equilibrium spin period) of $\rm
P_{0}\approx3.576$ ms. This value, combined with the surface magnetic
field ($\rm B\approx3.145\times10^{8}$ G) and assuming a NS with a
radius of 10 km and a canonical mass of $\rm 1.4 \ \Msun$, can be used
to infer the typical accretion rate that reaccelerated the NS during
the low mass X-ray binary phase. By using equation (8) of
\citet{vandenheuvel09}, we find that the system past accretion history
likely proceeded at a sub-Eddington rate
$(\dot{M}/\dot{M}_{EDD}\sim0.02)$, as expected from the typical
  evolution of close binary systems with light donor stars
  \citep{tauris99,istrate14}. Although the mass accretion rate
strongly depends on the NS radius, the general result does not change
assuming different radii or even different NS masses.

\subsection{Possible candidate companion stars}
\label{comH}

As can be seen from Figures \ref{cmd}, \ref{cmdwd} and \ref{mp}, the
possible companion to 47TucH appears to have properties quite
different from those observed for the other companions, first of
  all its much larger distance from the MSP nominal position
  ($0.24\arcsec$), which corresponds to almost twice our astrometric
  uncertainty. Moreover, following the procedure adopted in the
previous section, we derived for this object a mass of $\rm
0.37\pm0.05 \ \Msun$. This value, combined with the binary system
total mass of $\rm 1.61 \ \Msun$ \citep[][]{freire03}, would imply a
PSR mass of $\rm \sim1.25 \ \Msun$, a value slightly lower than
expected for a recycled PSR, although still acceptable within the
uncertainties. Its position in the CMD is compatible also with
  the CO WD cooling sequence, which would increase the probability of
  a chance coincidence to $\sim 2-3\%$.  Furthermore, at odds with the
  others objects, this candidate counterpart occupies an anomalous
  region in the orbital period companion mass plane shown in
  Figure~\ref{mp}. Although this anomaly could be real (since 47TucH
  has a large eccentricity, probably due to some kind of dynamical
  interaction), all these pieces of evidence suggest that the observed
  object is probably an isolated WD and the true companion star is
  still under the detection threshold (see
    Section~\ref{nondec}).

A possible candidate companion to MSP 47TucI has been also
  detected (see Figure~\ref{fig1} for the finding chart). This is a
  binary system with a short orbital period ($\sim0.23$ days) and a
  very small eccentricity. From the PSR mass function, the companion
  is expected to be a very low mass star ($\rm M_{COM}\sim0.015
  \ \Msun$). The absence of radio eclipses, probably due to a low
  inclination angle, prevents its characterization as a black widow
  system. At $0.15\arcsec$ from the PSR position, we identified a star
  located at the faint-end of the cluster main sequence (see
  Figure~\ref{cmd} and Table~\ref{tab2}). If we assume that the
  companion is a bloated star seen in a binary system with a low
  inclination angle, such a CMD position could be reasonable.
  However, the lack of any significant variability related to the
  orbital period prevents us from firmly associating this candidate to
  the MSP. In fact, the orbital period coverage of the F390W images is
  too poor, while the signal to noise ratio of the F300X data allows
  us to only infer that, in case of photometric variability, the
  maximum variation amplitude must be smaller than $\sim0.8$ mag.  We
  therefore conclude that it is more likely that the real companion
  star is still under the detection threshold. Indeed, the
probability of a chance coincidence with a main sequence star is non
negligible ($\rm \sim45-50\%$). We finally note that another object
lies within the astrometric uncertainty circle, but its association
with the MSP can be excluded, since it is a common CO WD, with
properties incompatible with the MSP timing ephemeris.}

\subsection{Non detections}
\label{nondec}
No interesting counterparts have been identified for all the other
known binary MSPs. These non-detections are likely due to companion
stars still under the detection threshold (as in the case, e.g., of
47TucE and the black widow 47TucJ), or to the severe crowding
conditions of the area surrounding the MSP positions (as in the case
of the black widows 47TucO and 47TucR). No search could be performed
for 47TucX since its position is outside the field of view.

 Considering that the companion to 47TucJ should be a
  non-degenerate object, its non-detection in UV passbands cannot be
  used to get useful information on its properties. Instead, the
  counterpart to 47TucE is expected to be He WD, which remains
  undetected down to our limiting magnitudes ($\sim25$ in the F300X
  filter and $\sim25.5$ in the F390W filter).  Hence, taking into
  account that the cooling age of a $\sim 0.17 \Msun$ WD at these
  detection thresholds is larger than the cluster age ($\sim10-11$
  Gyr; \citealp{gratton03,hansen13}), it is unlikely that this star
  has a mass similar to that estimated for the other companions.  It
  is more probable that it is more massive than $0.2\Msun$
  (corresponding to a faster cooling) and its cooling age is larger
  than 1 Gyr.  The same should apply also to the case of 47TucH if its
  true companion is still under our detection limits (as suggested
  above).  Interestingly, according to the theoretical relation of
  \citet{tauris99} and the orbital periods of 47TucE and 47TucH
  ($\sim2.3$ and $\sim2.4$ days, respectively), the companions to both
  these MPSs are indeed expected to have masses $\gtrsim0.2\Msun$.

\subsection{The companion to the redback 47TucW}
\label{w}
47TucW is the only redback identified, so far, in 47 Tucanae. It
is a binary MSP with a spin period of 2.35 ms, an orbital period of
$\sim3.2$ hr \citep{camilo00} and a companion mass of $\rm \sim0.15
\ \Msun$. The first optical identification of this system was
presented in \citet{edmonds02}, who suggested that the companion is a
perturbed and non degenerate star with a light curve structure
indicating a strong heating by the PSR flux. In Figure~\ref{curva} we
show, for both the filters, the light curves we obtained by folding
our photometric measurements with the most updated radio timing
ephemeris (Freire et al., in preparation). The zero orbital phase has been set at
the PSR ascending node time\footnote{Please note that we are using a
  different formalism with respect to \citet{edmonds02}.}. As can be
seen, in agreement with previous works \citep{edmonds02, bogdanov05}, the light curves present a
single maximum-minimum structure, likely due to the heating by the PSR
flux. Unfortunately, the star has been measured above the detection
threshold only near its maximum luminosity. Nonetheless,
modeling\footnote{We used the ``Graphical Analyzer for TIme Series'',
  a software aimed at studying stellar variability phenomena,
  developed by Paolo Montegriffo at INAF-Osservatorio Astronomico di
  Bologna.}  the sinusoidal light curve, we found that, in both the
filters, the companion spans $\sim3.5$ magnitudes between the maximum
and the derived minimum, in agreement with previous observations. The
best fit-model is shown as a solid curve in
Figure~\ref{curva}. Interestingly, the light curve structure is more
similar to the ones observed for black widow than for redback
companions, which usually, but not always, show a double minimum-maximum structure due
to tidal deformation \citep[see
  e.g.][]{fer03_msp,cocozza08,pallanca10,li14}. The CMD position of
the companion during the maximum and at a mean phase (as derived by
the adopted model) is shown in Figure~\ref{cmd}. The system is located
between the main sequence and the WD cooling sequence, where no normal
stars are expected and thus suggesting a perturbed and strongly heated
companion star. Again, at odds with other redback systems, this lies
in a region more similar to that occupied by the two black widow
companions identified so far in GCs \citep[][Cadelano et al.,
  2015]{pallanca14b}.  The X-ray counterpart to 47TucW shows a
variability which is likely due to an intra-binary shock between the
PSR wind and the matter lost by the companion
\citep{bogdanov05}. Interestingly, as discussed by \citet{bogdanov06}, the minimum of the X-ray light
curve is displaced with respect to the optical one. Such a behavior
has been also noticed for the black widow M71A \citep{cadelano15},
thus further strengthening the connection of this MSP with black widow
systems. All this allow us to speculate that a scenario where 47TucW will evolve
into a canonical MSP with a He WD companion \citep[as in the case of
  MSP-A in NGC 6397; see][]{burderi02} is somewhat unlikely, opening the possibility to an
evolution toward the black widow stages. Indeed
such an evolutionary path has been already suggested by the
simulations of \citet{benvenuto14}. The identification of new redback
companions will shed light on this possibility.

\section{SUMMARY}
By using ultra-deep, high resolution UV WFC3/HST observations of 47
Tucanae, we identified the companions to four binary MSPs (47TucQ,
47TucS, 47TucT and 47TucY) and confirmed the two already known objects
(COM-47TucU and COM-47TucW). The optical counterparts have coordinates
compatible, within the errors, with the PSR nominal positions. In the
CMD, all the objects are located in the He WD cooling sequence, as
expected from the MSP canonical evolutionary scenario. The only
exception is the companion to the redback system 47TucW, which is
located in an anomalous region between the main sequence and the WD
cooling sequence, suggesting that it is a low-mass MS star
highly perturbed and heated by the PSR flux. We compared the observed
CMD positions of the detected He WD companions with a set of cooling
tracks and derived the companion main properties (as masses, cooling
ages, temperatures) and also some constraints on the PSR masses. All
the companion stars have masses between $\rm \sim 0.15 \ \Msun$ and
$\rm \sim 0.20 \ \Msun$, and all the derived cooling ages are
  smaller than the cluster stellar population age. The orbital
periods vs companion masses are in fair agreement with the
evolutionary models of \citet{tauris99} and \citet{istrate14}.  By
combining the cooling age with the PSR spin down rate we found that
the accretion history of 47TucU likely proceeded at a sub-Eddington
rate.

 By taking into account our astrometric uncertainty
  ($0.15\arcsec$), we also detected a star having a position
  marginally compatible with that of 47TucH. However, its photometric
properties would imply a PSR mass lighter than expected for a recycled
NS. Moreover, its position in the plane of orbital period
vs. companion mass is in clear disagreement with the theoretical
predictions. While this could be due to its high eccentricity, the
object could be just a chance coincidence and further investigations
are needed before confirming its association to 47TucH. A possibile
counterpart to 47TucI has been also identified in a star located in a
low luminosity region of the cluster main sequence. However its
distance from the MSP position ($0.15\arcsec$) and the absence of any
optical variability related to the orbital period do not allow us to
asses a clear connection with the binary system.

Finally we discussed how the properties of COM-47TucW are more similar
to those usually observed for black widows than for redbacks, thus
opening the possibility that this MSP could be the prototype of a
redback evolving into a black widow system.

\section*{Acknowledgements}
This research is part of the project {\it Cosmic-Lab}
(\url{http://www.cosmic-lab.eu}) funded by the European Research
Council under contract ERC-2010-AdG-267675.  The authors kindly thank
P. Bergeron and S. Cassisi for the help with the cooling tracks and
isochrones, and T. Tauris and A. Istrate for providing us with their
simulation data. M.C. thanks A. Istrate for the useful discussion.

\clearpage

\begin{deluxetable}{cccccccc}
\tablecolumns{8}
\tablewidth{0pt}
\tablecaption{Radio timing ephemeris of the analyzed MSPs}
\tablecomments{From left to right: MSP name, position, offset from the
  GC center, orbital period, mass function and
  characteristic age. Numbers in parentheses are uncertainties in the last digits quoted. Reference: \citet{freire03}. }
\tablehead{\colhead{MSP} & \colhead{$\alpha$ (h m s)} & \colhead{$\delta$ ($^{\circ}$ $\arcmin$ $\arcsec$)} & \colhead{Offset ($\arcmin$)} &  \colhead{$\rm P_{ORB}$ (d)} &   \colhead{$f$ ($\Msun$)}  & \colhead{$\tau_{age}$ (Gyrs)$^{b}$}}  
\startdata
47TucH  & 00 24 6.7014(3) & -72 04 6.795(1) & 0.77 &  2.36 &   1.927$\times10^{-3}$  & $>0.93$ \\
\hline
47TucI & 00 24 7.9330(3) & -72 04 39.669(1) & 0.29 & 0.23 & 1.156$\times10^{-6}$ & $>0.23$ \\
\hline 
47TucQ  & 00 24 16.4891(4) & -72 04 25.153(2) & 0.98 & 1.19  &  2.374$\times10^{-3}$  & $>1.43$ \\
\hline
47TucS  & 00 24 3.9779(4) & -72 04 42.342(1) & 0.19  & 1.20  &   3.345$\times10^{-4}$  & $>0.91$  \\
\hline
47TucT  & 00 24 8.548(2) & -72 04 38.926(7) & 0.34 & 1.13  &   2.030$\times10^{-3}$  & $>0.32$ \\
\hline 
47TucU & 00 24 9.8351(2) & -72 03 59.6760(9) & 0.94  & 0.43  &   8.532$\times10^{-4}$  & $2.5$ \\
\hline
47TucW$^{b}$ & 00 24  6.059(1) & -72 04 49.084(2) & 0.08$^{a}$  & 0.13$^{a}$  &  8.77$\times10^{-4}$ & $>1.15$  \\
\hline
47TucY$^{b}$ & 00 24 1.4023(3) & -72 04 41.837(1) & 0.37  & 0.52$^{a}$ &  1.195$\times10^{-3}$  & $>2.2$  \\
\hline
\enddata
\tablenotetext{a}{\url{http://www.naic.edu/~pfreire/GCpsr.html.}}
\tablenotetext{b}{P. Freire et al. 2015, in preparation.}        
\label{tab1}                      
\end{deluxetable}

\begin{deluxetable}{lccccc}
\tablecolumns{6}
\tablewidth{0pt}
\tablecaption{Optical properties of the companion stars}
\tablecomments{From left to right: MSP name, position, distance from
  the radio MSP nominal position, F300X and F390W magnitudes and the
  relatives uncertainties.}
\tablehead{\colhead{Name} & \colhead{$\alpha$ (h m s)} & \colhead{$\delta$ ($^{\circ}$ $\arcmin$ $\arcsec$)} & \colhead{dist ($\arcsec$)} & \colhead{$\rm m_{F300X}$}  & \colhead{$\rm m_{F390W}$}}  
\startdata
COM-47TucQ  & 00 24 16.489 &  -72 04 25.209   &   0.04   &$23.19\pm0.02$ & $23.63\pm0.05$ \\
\hline
COM-47TucS  & 00 24 3.977 & -72 04 42.385    &   0.03    &$23.29\pm0.02$ & $23.80\pm0.05$\\
\hline
COM-47TucT  & 00 24 8.549 &  -72 04 38.965   &   0.04    &$23.07\pm0.02$ & $23.56\pm0.03$\\
\hline 
COM-47TucU & 00 24 9.835 & -72 03 59.746   &   0.06   &$20.40\pm0.01$ & $20.85\pm0.03$\\
\hline
COM-47TucW & 00 24 6.063 & -72 04 49.133    &   0.06    & 24.28$^{a}$ & 23.62$^{a}$  \\
\hline
COM-47TucY & 00 24 1.401 & -72 04 41.875    &   0.04    & $22.16\pm0.02$ & $22.69\pm0.04$\\
\hline
COM-47TucH?  & 00 24 6.755    & -72 04 6.781    &   0.24    & $23.39\pm0.02$ & $24.25\pm0.05$ \\
\hline    
COM-47TucI? & 00 24 7.953 & -72 04 39.559 & 0.15 & $24.14\pm0.04$ & $22.43\pm0.03$\\
\hline
\enddata
\tablenotetext{a}{The values for COM-47TucW correspond to the mean
  magnitudes of the best-fit models (see Figure~\ref{curva}).}
\label{tab2}    
\end{deluxetable}

\begin{deluxetable}{lcccccc}
\tablecolumns{6}
\tablewidth{0pt}
\tablecaption{Derived properties of the five MSPs with He WD companions}
\tablecomments{From top to bottom: companion mass, age, temperature,
  luminosity, PSR mass and inclination angle. }
\tablehead{\colhead{Parameter} & \colhead{47TucQ} & \colhead{47TucS} &
  \colhead{47TucT} & \colhead{47TucU} & \colhead{47TucY} }
\startdata
$\rm M_{COM} \ (\Msun)$ & $\sim0.15$ & $0.17^{+0.03}_{-0.02}$ & $0.16^{+0.025}_{-0.01}$ & $0.171^{+0.002}_{-0.003}$ & $0.17\pm0.02$  \\ \hline
Age (Gyrs) & $\sim5.5$ & $6.4^{+1.7}_{-6.0}$ & $5.1^{+0.9}_{-3.5}$ & $0.88^{+0.05}_{-0.06}$ & $2.2^{+1.0}_{-1.6}$  \\ \hline
T ($10^{3}$ K) &  $\sim7.6$ & $8.1^{+1.0}_{-0.7}$ & $8.0^{+0.6}_{-0.5}$ & $11.9^{+0.2}_{-0.5}$ & $9.6^{+0.5}_{-1.2}$  \\ \hline
L ($\rm 10^{-3} \ L_{\odot}$)  &  $\sim9.5$ & $8.1^{+1.0}_{-0.3}$ & $10.0^{+0.5}_{-0.6}$ & $158^{+7}_{-17}$ & $23.0\pm5$  \\ \hline
$\rm M_{PSR} \ (\Msun)$ & $<1.57$ & $<4.69$ & $<1.58$ & $<2.30$ & $<2.22$  \\ \hline
$i \ (^{\circ})$ & $>58$ & $>26$ & $>57$ & $>42$ & $>45$   \\ \hline  
\enddata
\label{tab3}
\end{deluxetable}

\begin{figure*}
\begin{center}
\leavevmode
\includegraphics[width=7cm]{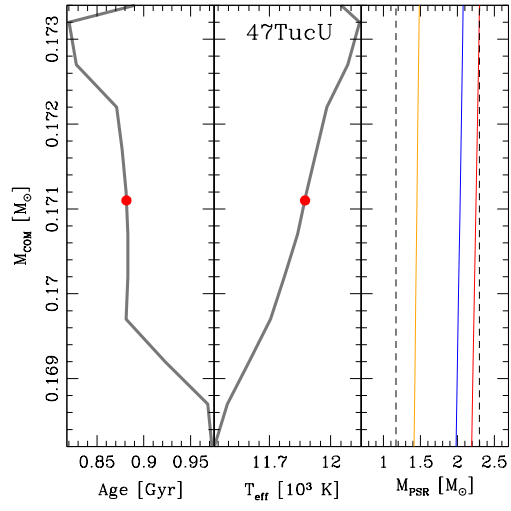}
\caption{As in Fig. \ref{bo1}, but for 47TucU.}
\label{bo2}
\end{center}
\end{figure*}

\begin{figure*}
\begin{center}
\leavevmode \includegraphics[width=13cm]{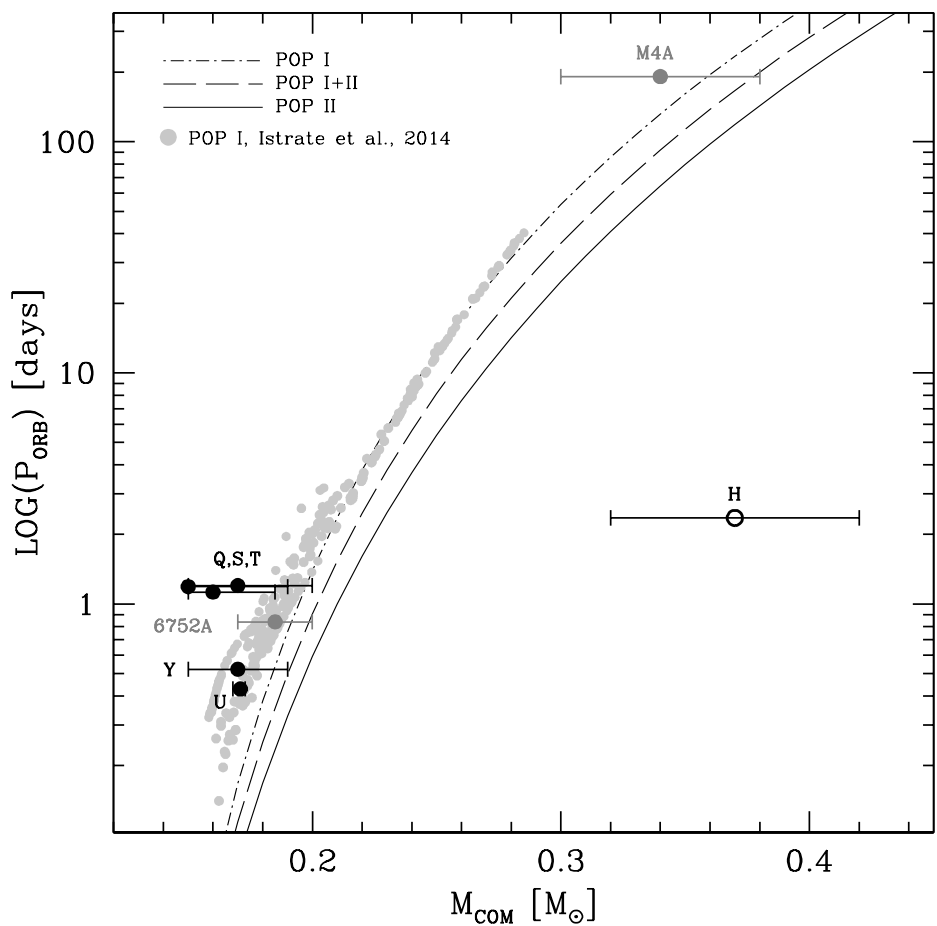}
  \caption{MSP orbital periods plotted as a function of the best-fit
    companion masses, for each identified object (see lables), plus
    the ones detected in NGC 6752 and M4 \citep[dark gray points;
      see][]{fer03_msp,sigurdsson03}. The three curves correspond to
    the theoretical predictions of \citet{tauris99} for three
    different stellar population progenitors, as reported in the
    top-left legend. The light gray points correspond to the
    theoretical results obtained by \citet{istrate14}.}
  \label{mp}
\end{center}
\end{figure*}

\begin{figure*}
\begin{center}
\leavevmode
\includegraphics[width=13cm]{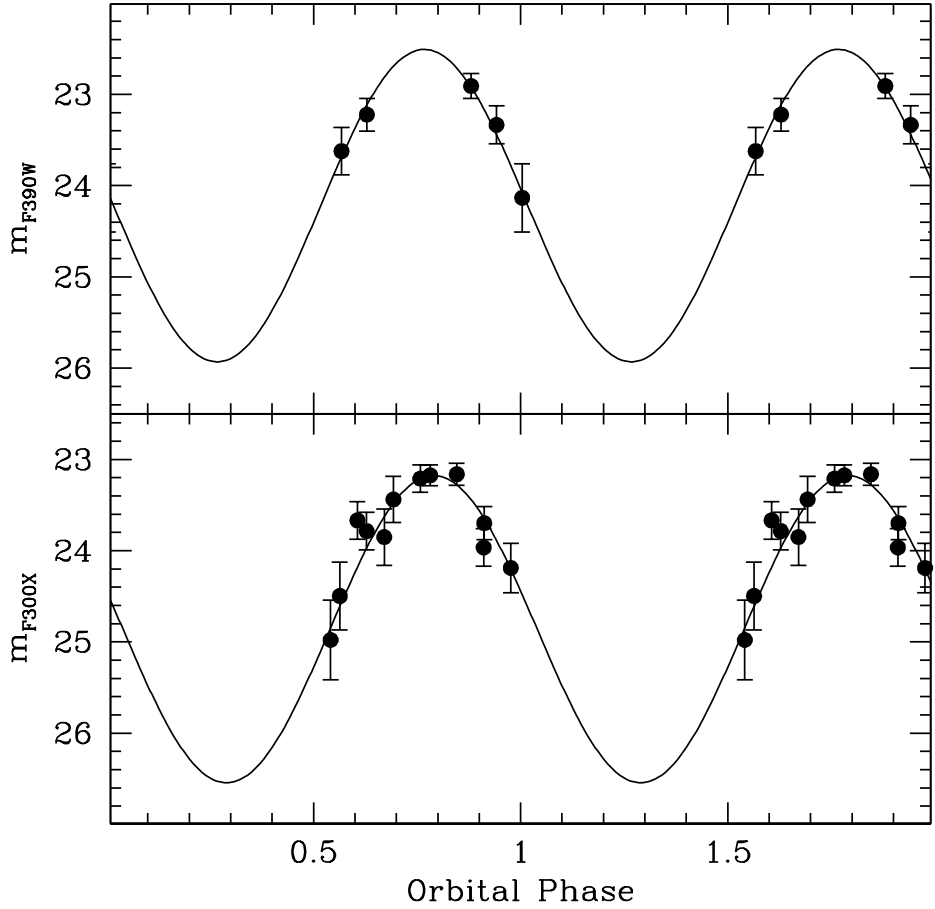}\\
  \caption{Light curves of COM-47TucW in the F390W (upper panel) and
    F300X (lower panel). The two curves are folded with the radio
    parameters and two periods are shown for clarity.  The black curve
    in each panel is the best analytical model obtained independently
    for each filter.}
  \label{curva}
\end{center}
\end{figure*}

\end{document}